# How long can LBVs sleep? A long-term photometric vaiability and spectral study of the Galactic candidate luminous blue variable MN 112

O. V. Maryeva[1,2]⋆, S. V. Karpov[3,2], A. Y. Kniazev[4,5,6], V. V. Gvaramadze[6]

[1] *Astronomical Institute of the Czech Academy of Sciences, Fričova 298, 25165 Ondřejov, Czech Republic*
[2] *Special Astrophysical Observatory of the Russian Academy of Sciences, Nizhnii Arkhyz, 369167, Russia*
[3] *Institute of Physics of the Czech Academy of Sciences, CZ-182 21 Prague 8, Czech Republic;*
[4] *South African Astronomical Observatory, PO Box 9, 7935 Observatory, Cape Town, South Africa*
[5] *Southern African Large Telescope Foundation, PO Box 9, 7935 Observatory, Cape Town, South Africa*
[6] *Sternberg Astronomical Institute, Moscow State University, Universitetsky pr., 13, Moscow, 119992, Russia*

**ABSTRACT**

Luminous Blue Variables (LBVs) are massive stars that show strong spectral and photometric variability. The question of what evolutionary stages they represent and what exactly drives their instability is still open, and thus is it important to understand whether LBVs without significant ongoing activity exist, and for how long such dormant LBVs may "sleep". In this article we investigate the long-term variability properties of the LBV candidate MN 112, by combining its optical and infrared spectral data covering 12 years with photometric data covering nearly a century acquired by both modern time-domain sky surveys and historical photographic plates. We analyze the spectra, derive physical properties of the star by modelling its atmosphere and use a new distance estimate from *Gaia* data release 3 (DR3) to determine the position of MN 112 both inside the Galaxy and in the Hertzsprung–Russell diagram. Distance estimation increased in almost two times in comparison with *Gaia* DR2. Due to that MN 112 moved in upper part of the diagram and according to our modeling it lies on evolutionary track for star with initial mass $M_* = 70\,M_\odot$ near Humphreys–Davidson limit. Given the absence of any significant variability we conclude that the star is a dormant LBV that has been inactive for at least a century now.

## 1 INTRODUCTION

Luminous Blue Variables (LBVs) are massive stars on an important phase of their lifetime when their luminosity approaches the level that starts influencing their overall stability, and the stars reach the region close to empirical Humphreys–Davidson limit (Humphreys & Davidson 1994; Vink 2012; Weis & Bomans 2020). During this phase, the stars lose a large fraction of their initial masses through strong stellar winds and occasional giant eruptive events, and thus LBV phase is essential in defining consecutive stellar evolution and the properties of pre-supernova stars. According to theoretical models, for isolated massive stars with masses above $40\,M_\odot$, LBV is an intermediate evolutionary phase before moving towards Wolf-Rayet (WR) stars (Conti 1984; Groh et al. 2014), while for lower-mass stars (∼25-40 $M_\odot$) this phase is probably the final stage of stellar evolution. Low luminosity LBVs could be progenitors of core-collapse supernovae (SNe) of IIL/b types (Groh et al. 2013; Moriya et al. 2013). Alternatively, LBVs may be the mass gainer products of binary evolution as suggested by Smith & Tombleson (2015); Smith (2016, 2019) and further critically discussed by Humphreys et al. (2016); Davidson et al. (2016). Massive LBVs may also be direct SNe progenitors if they are the products of binary mergers (Justham et al. 2014).

Main observational characteristics of LBVs is dramatic spectroscopic and photometric variability. LBVs show both short timescale stochastic variations of brightness with amplitude up to a few tenths of magnitude (Humphreys & Davidson 1994) and a long-term variability – so called S Doradus (S Dor) cycle (van Genderen 2001; Vink 2012). Typically S Dor cycles last for several years and their amplitude is 1–2 mags (Humphreys et al. 2016). During S Dor cycles, photometric variability is always accompanied by changes in effective temperature, which causes LBVs to shift on the Hertzsprung–Russell (H–R) diagram.

For understanding the nature of LBV phenomenon; the fraction of binaries among LBVs; their interaction with the environment; etc., we need a large number of objects reliably classified as LBVs. Different methods are presently being used to search for new LBV candidates, such as: selection of targets through the detection of their mid-infrared circumstellar nebulae in Spitzer and WISE archives (e.g. Gvaramadze et al. (2010b)); selection based on colour–colour diagrams (e.g. Massey et al. (2016) or Kraus (2019)); search for objects with strong emission in the H$\alpha$ line (Valeev et al. 2010).

On the other hand, as LBVs are rare, inclusion of even one impostor into their lists may significantly affect the statistical conclusions on their evolutionary status (see for example discussion in Humphreys et al. (2014, 2016, 2017)). Thus, further confirmation of LBV candidates is an important task. Apart from basic LBV characteristics – spectral and photometric variabiliy – they should also have low wind velocities and lack of warm circumstellar dust (Humphreys et al. 2017). As the S Dor phase is associated with variable stellar winds, the confirmation of LBVs may in principle be done based on high resolution snap-shot spectra, as suggested by Groh & Vink (2011). However, that is problematic for the majority of LBV candidates due to their faintness (due to high extinction for Galactic ones, and large distances – for extragalactic objects), and realistically only long-term photometric and spectroscopic monitoring may confirm that they are *bona fide* LBVs. It makes confirmation of LBV status a complex problem, and forces us to return to specific LBV candidates again and again.

⋆ E-mail: olga.maryeva@asu.cas.cz

**Table 1.** Summary of spectral data used in present work.

| Date | Telescope | Instrument | Sp. range (nm) | Resolution (Å) | Exposure (sec) | Pub. |
|---|---|---|---|---|---|---|
| | | | Optical Range | | | |
| 2009 May 5 | Calar Alto | TWIN | 350-560 / 530-760 | 3000 | 600 | Paper I / Paper I |
| 2009 June 20 | Russian 6-m | SCORPIO | 405-580 | 1000 | 1800 | Paper I |
| 2012 July 13 | Calar Alto | TWIN | 350-560 / 530-760 | 3000 | 600 | |
| 2015 August 16 | Russian 6-m | SCORPIO | 400-790 | 500 | 1160 | |
| 2019 August 13 | Tartu AZT-12 | ASP-32 | 400-730 | 1000 | 3600 | Cochetti et al. (2020) |
| 2021 April 13 | Russian 6-m | SCORPIO | 565-740 | 1000 | 2280 | |
| 2021 June 08 | Perek's 2-m | Coude | 625-674 | 13000 | 4800 | |
| | | | Near-IR Range | | | |
| 2010 June 20 | IRTF | SpeX | 800-2400 | 2000 | 8 × 50 | K-band published in Stringfellow et al. (2012) |
| 2013 July 07 | Gemini | GNIRS | 2290-2480 | 6000 | 8 × 192 | Cochetti et al. (2020) |

Galactic star MN 112[1] was known as object with emission-line spectrum since mid-70s (Dolidze 1975). However the interest to the star appeared in 2009 when MN 112 was identified as new Galactic LBV candidate via detection of its circumstellar nebula (Gvaramadze et al. (2010b), Gvaramadze et al. (2010a) hereafter Paper I). Its nebula is visible in infrared (IR) range and it has clear spherically-symmetric shape (Paper I). In the optical part of the MN 112's spectrum there are lines of H, He I, Fe III, N II and Si II. The spectrum as a whole looks similar to the one of classical LBV star P Cygni (Paper I). Numerical modeling of MN 112's atmosphere done by Kostenkov et al. (2020) shows that both stars MN 112 and P Cygni also have similar values of luminosities. Cochetti et al. (2020) investigated MN 112 based on K-band spectrum and its location in the colour–colour diagram. Cochetti et al. (2020) concluded that the star shows characteristics typically seen in LBV stars: the Mg II lines are more intense than the adjacent Pfund lines; and in *JHK* colour-colour diagram MN 112 lies in region of interstellar reddened LBVs.

In this paper we study long-term spectral and photometric variability of MN 112 and discuss its evolutionary status. We present collected spectral and photometric observations in Section 2, and then analyze these observations in Section 3.1. In Section 3.2 we present the results of numerical modeling of MN 112's atmosphere.

In Section 4 we discuss the properties of MN 112 as a whole and conclude our study in Section 5.

## 2 OBSERVATIONS

### 2.1 Spectroscopy

To investigate spectral properties and variability of MN 112, we combined archival data including both already published and previously unpublished spectra, with two new low and medium resolution spectra that we acquired in 2021. The details of used spectral data are given in Table 1.

The medium resolution spectrum ($R \simeq 13000$) was obtained using the coude spectrograph attached to the Perek's 2-m telescope at the Ondřejov observatory of the Astronomical Institute of the Czech Academy of Sciences (Slechta & Skoda 2002). Spectrum covers the 6250–6740 ÅÅ range and includes $H\alpha$ and He I$\lambda$6678 lines. The reduction of the Ondřejov data was performed using IDL-based package, which includes all standard steps – bias subtraction, flatfielding, wavelength calibration using spectrum of ThAr calibration lamp.

Low resolution spectrum ($R \simeq 1000$) was obtained with the Russian 6-m telescope with the Spectral Camera with Optical Reducer for Photometric and Interferometric Observations (SCORPIO) (Afanasiev & Moiseev 2005) with the VPHG 1200R grism, which provided a spectral range of 5650-7350 ÅÅ. The spectrophotometric standard star BD+33 2642 was observed for flux-calibration on the same night. Moreover in the General Observational Data Archive of Russian 6-m telescope[2] we found two spectra also obtained with SCORPIO in June 2009[3] and in August 2015. All SCORPIO spectra were reduced in a uniform way using the ScoRe package[4] specifically created for the SCORPIO long-slit spectroscopic data reduction.

The near-IR spectra were taken with the cross-disperser medium-resolution spectrograph SpeX (Rayner et al. 2003) mounted at the 3.0-m NASA Infrared Telescope Facility (IRTF) on Mauna Kea,

---

[1] [GKF2010] MN 112 with coordinates $\alpha$=19:44:37.60, $\delta$=+24:19:05.74.

[2] https://www.sao.ru/oasis/cgi-bin/fetch?lang=en
[3] The spectrum obtained at June 2009 was published in Paper 1.
[4] ScoRe package is available at http://www.sao.ru/hq/ssl/maryeva/score.htm.

Hawaii. The observations were carried out with two different set-ups with a resolving power $R \simeq 2000$ at 0.8-2.4 $\mu$m. Several spectral orders were simultaneously recorded during a single exposure with significant wavelength overlap between the adjacent orders making it easier to preserve the continuum shape. Standard techniques for near-IR spectroscopy reduction were used with the software Spextool (Cushing et al. 2004) dedicated to reducing data obtained with SpeX. K-band spectrum from this set of observations was published by Stringfellow et al. (2012).

We included in our analysis two spectra obtained with the Cassegrain Twin Spectrograph (TWIN) of the 3.5-m telescope in the Observatory of Calar Alto, Spain. One of them was published in Paper I. Also Dr. Michaela Kraus kindly provided us with data which were published by Cochetti et al. (2020): K-band spectrum from GEMINI and optical one from Tartu 1.5-m telescope.

### 2.2 Photometry

In order to characterize long-term photometric behaviour of MN 112 we combined its light curve from a diverse data from various instruments and archives. As most of them are acquired in vastly different photometric passbands, we converted them all to a common photometric system – Pan-STARRS $g$ band – using simple assumptions about colour stability of the object. Below we will give the details on individual data sets we used, and on the specifics of their cross-calibration. The data are also summarized in Table 2.

#### 2.2.1 Pan-STARRS

We acquired time-resolved photometric data for MN 112 from Pan-STARRS DR2 catalogue table containing information on individual detections of the source [5] during Pan-STARRS PS1 $3\pi$ Steradian Survey (Chambers et al. 2016). As the source is sufficiently bright, it is often saturated on individual Pan-STARRS exposures. To reject these bad measurements we only selected the entries with psfQfPerfect > 0.95 (i.e. with less than 5% of masked pixels inside PSF), and used the fluxes measured by PSF fitting to derive AB magnitudes and their errors. We got 14 photometric points in Pan-STARRS $g$ filter, and 6 – in $r$ one[6]. In all other filters no reliable data points were found.

Both of $g$ and $r$ band measurements show variations with amplitude greater than 0.1 magnitudes. However, none of the $g$ band measurements are acquired on the same night as $r$ band ones, making direct study of fine $g - r$ colour variation impossible due to rapid low-amplitude variability of the source brightness. However, as its rough estimate, we may use the difference of mean fluxes (with corresponding errors of the means) in these bands as

$$g - r = (15.67 \pm 0.01) - (13.37 \pm 0.01) = 2.30 \pm 0.02 \quad (1)$$

#### 2.2.2 Zwicky Transient Facility

The field of MN 112 has been repeatedly observed by 48-inch Samuel Oschin robotic telescope of Palomar Observatory during Zwicky Transient Facility (ZTF) (Bellm et al. 2019) survey. We downloaded

---

[5] Available online in Catalogs.MAST at https://catalogs.mast.stsci.edu/panstarrs/
[6] While Pan-STARRS filters response is not identical to SDSS one, for simplicity we will use through the text the names $g$ and $r$ solely for measurements in Pan-STARRS filters.

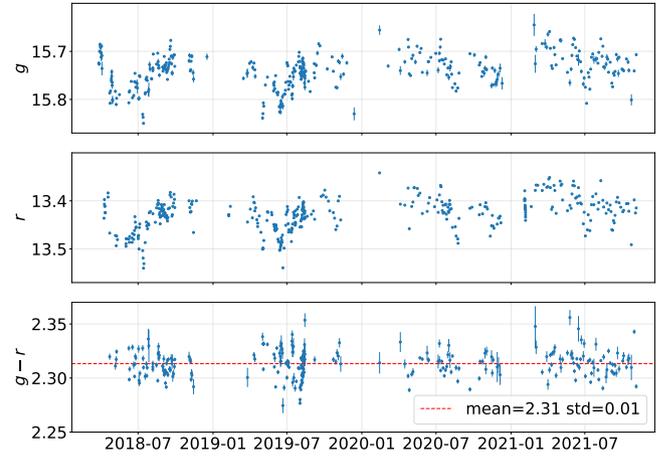

**Figure 1.** Light curves of MN 112 in Pan-STARRS $g$ (upper panel) and $r$ (middle panel) measured in ZTF DR9 data. Lower panel also shows $g - r$ colour estimated using consecutive measurements in corresponding filters during the same observational night.

798 survey images in ZTF_$g$ and ZTF_$r$ filters containing the object from ZTF DR9 data archive (Masci et al. 2019) at NASA/IPAC Infrared Science Archive[7].

We processed the images using custom pipeline based on STDPipe routines (Karpov 2021). Processing included SExtractor (Bertin & Arnouts 1996) aperture photometry for all field stars in circular apertures with radius equal to mean full width at half maximum (FWHM) of every image, with subtraction of local background estimated within circular annuli between radii with 5 and 7 FWHM sizes. Instrumental magnitudes have been then calibrated for every frame using Pan-STARRS DR1 catalogue measurements for field stars and photometric model including spatially varying zero point (linear term in $x$ and $y$), and $g - r$ colour term to accommodate for both per-frame atmospheric secondary extinction and systematic differences of ZTF_$g$ and ZTF_$r$ filter throughputs from Pan-STARRS $g$ and $r$ ones. We rejected the images where photometric calibration failed, or the object aperture contained masked pixels, or derived photometric accuracy has been worse than 0.3 magnitudes. Finally, we used mean value of $g - r$ colour of MN 112 from Eq. 1 to reduce the calibrated measurements of the object to Pan-STARRS photometric system. The colour terms derived for conversion between ZTF and Pan-STARRS filters are small (with mean values of -0.07 for ZTF_$g$ to $g$ conversion, and -0.11 – for ZTF_$r$ to $r$), even slight variations of the object colour with time should not significantly influence the results of such conversion.

Resulting light curves in Pan-STARRS $g$ and $r$ bands, along with $g - r$ colours estimated using the measurements from inside the same night (separated by no more than 0.3 days), are shown in Figure 1. The light curves show rapid aperiodic variability on various time scales, nicely correlated between both bands. The colour is essentially constant with mean value $g - r$=2.31, in good agreement with Eq. 1.

#### 2.2.3 APASS DR10

The American Association of Variable Star Observers (AAVSO) Photometric All-Sky Survey (APASS DR10) (Henden et al. 2018) con-

---

[7] Accessible at https://irsa.ipac.caltech.edu/Missions/ztf.html

**Table 2.** Summary of photometric data used in the study

|  | Start | End | Filter | N | Comments |
|---|---|---|---|---|---|
| Harvard plate collection | 1926.10.01 | 1970.08.26 | ATLAS_g | 12 | uniformly calibrated by DASCH |
| German photographic plates | 1962.07.23 | 1987.06.04 | V | 26 | uniformly calibrated by APPLAUSE |
| Moscow photographic plates | 1962.09.04 | 1994.09.05 | B | 10 | photometry is presented in Paper I |
| Russian 6-m photometry | 2009.06.21 | 2021.04.12 | B | 10 | |
|  |  |  | V | 22 | |
|  |  |  | R | 17 | |
| INT photometry | 2003.09.15 | 2011.08.30 | g | 1 | |
|  |  |  | r | 3 | |
| Pan-STARRS | 2009.06.20 | 2013.07.14 | g | 14 | |
|  |  |  | r | 6 | |
| APASS Epoch Photometry | 2011.11.18 | 2012.04.21 | APASS_B | 1 | |
|  |  |  | APASS_V | 4 | |
|  |  |  | APASS_g | 4 | |
|  |  |  | APASS_r | 4 | |
| ASAS-SN | 2015.02.11 | 2018.11.13 | APASS_V | 197 | |
|  | 2018.04.13 | 2022.02.14 | APASS_g | 303 | superseded by ZTF data |
| ZTF | 2018.03.26 | 2021.11.04 | ZTF_g | 363 | |
|  |  |  | ZTF_r | 384 | |
| FRAM-ORM photometry | 2021.05.08 | 2022.02.17 | B | 132 | |
|  |  |  | V | 272 | |
|  |  |  | R | 283 | |

tains several observations of MN 112 in its APASS Epoch Photometry Database[8]. In order to cross-calibrate them with Pan-STARRS photometric system we downloaded photometric information for all field stars within 30′ from the object from APASS DR10[9] and Pan-STARRS DR1[10] catalogues, cross-matched them, and derived the following set of approximate photometric conversions in this sky region for the redder objects with $g - r > 1.3$:

$$g = \text{APASS\_V} + 0.09 + 0.54 \cdot (g - r)$$
$$g = \text{APASS\_g} + 0.00 - 0.07 \cdot (g - r) \quad (2)$$
$$r = \text{APASS\_r} + 0.06 - 0.03 \cdot (g - r)$$

Using it and the colour from Eq. 1, we may regress all APASS_V, APASS_g and APASS_r measurements to Pan-STARRS $g$ band. Also, as the dataset contains APASS_B and APASS_V measurements on the same night, we may estimate object colour between these bands as:

$$\text{APASS\_B} - \text{APASS\_V} = 2.75 \pm 0.09 \quad (3)$$

### 2.2.4 ASAS-SN

The All-Sky Automated Survey for Supernovae (ASAS-SN, Shappee et al. (2014)) routinely monitors the whole sky using several stations each equipped with 14 cm aperture telephoto lenses and CCD cameras, providing ∼ 16″ angular resolution and ∼ 17 mag detection limit. We downloaded from its Light Curve Server (Kochanek et al. 2017) the differential[11] light curve extracted in a circular 16″ radius aperture with zero point calibrated to APASS DR9 catalogue zero point for all epochs available in the service. We then rejected all measurements with signal to noise (S/N) ratio worse than 10, and converted both APASS_V (before late 2018) and APASS_g (since early 2018) data points to Pan-STARRS $g$ magnitudes using photometric equations from Eq. 2 and colour from Eq. 1.

The resulting light curve shows some systematic shifts both in respect to ZTF data, and between simultaneous measurements of different ASAS-SN cameras. We attribute it to differences of reference (template image) flux measurements due to crowding, which is significant because of quite poor ASAS-SN angular resolution. Thus, we adjusted the data corresponding to earlier ASAS-SN measurements done in APASS_V filter by making it brighter by 0.25 magnitudes to match simultaneous ZTF points, and ignored later measurements done in APASS_g filter as they are completely superseded by ZTF data of better quality.

### 2.2.5 Harvard plate collection

Digital Access to a Sky Century @ Harvard (DASCH) project (Grindlay et al. 2012) is aimed towards digitization of the Harvard Astronomical Plate Collection, and providing an access for uniformly calibrated photometric data from these plates. We extracted from it the measurements of MN 112 calibrated to Pan-STARRS $g$ magnitudes using the ATLAS All-Sky Stellar Reference Catalog (ATLAS-REFCAT2), Tonry et al. (2018)). We decided to use much less restrictive quality cuts than Schaefer (2016) and only rejected the measurements with AFLAGS>524288 and one sigma error bars exceeding 1 magnitude. The rest of measurements do not show any significant variations over the span of more than 40 years.

---

[8] Accessible online as a link from AAVSO Variable Star Index (VSX) at https://www.aavso.org/vsx/index.php?
[9] Accessible online at https://www.aavso.org/download-apass-data
[10] Accessible online at VizieR(Ochsenbein et al. 2000)
[11] This kind of ASAS-SN light curve is produced by co-adding all images from individual epoch, subtracting template image, performing the photometry on differential image and then adding back the flux of object on template image. Due to co-adding, it provides the measurements of better quality than the photometry of individual images.

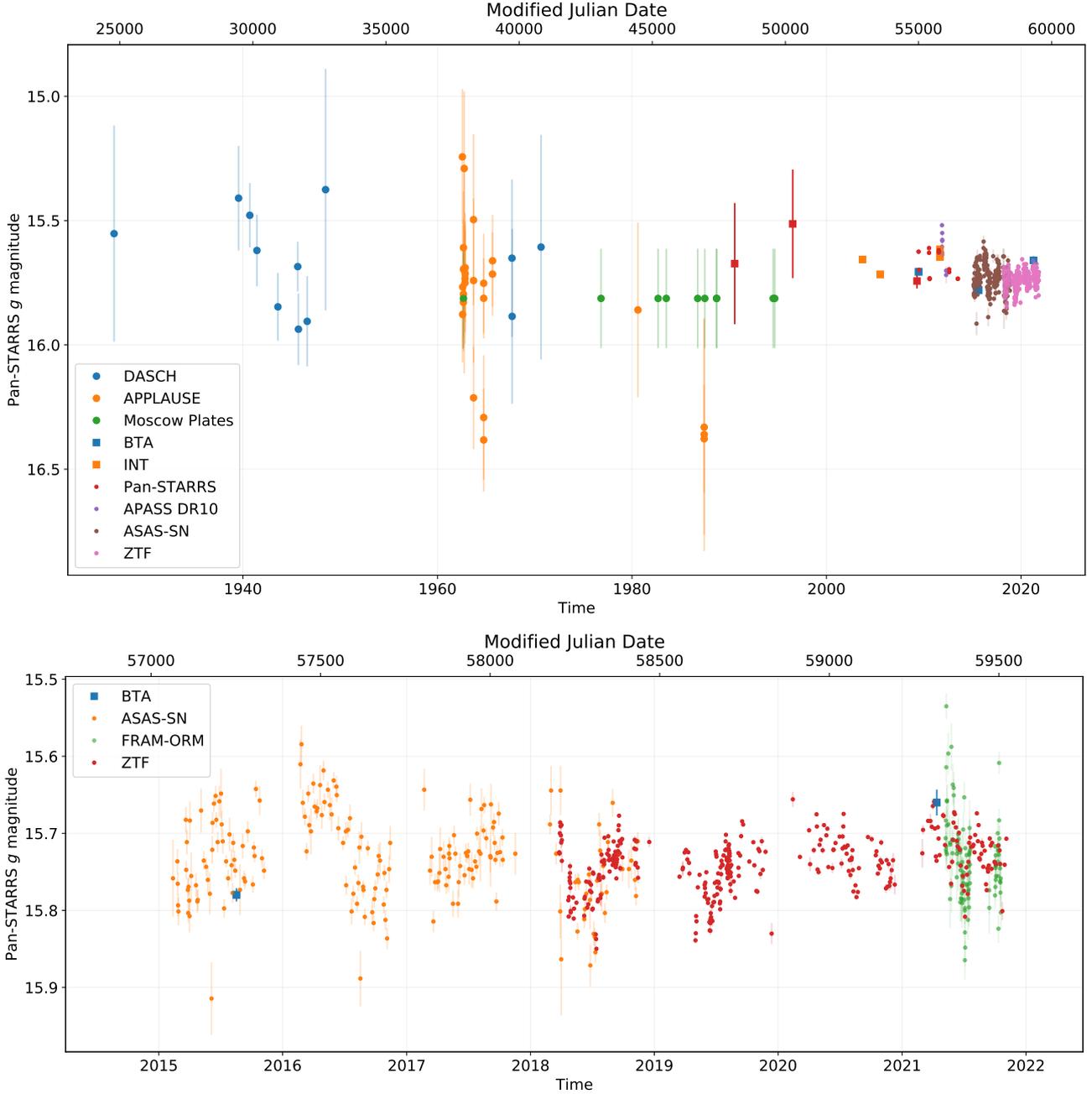

**Figure 2.** Light curve of MN 112 in Pan-STARRS *g* band assembled from various data sources. Upper panel shows the whole ∼100 years of data, while lower one – only last 7 years properly sampled by modern all-sky time-domain surveys.

*2.2.6 German photographic plates*

The plates from Potsdam, Hamburg and Bamberg plate archives have been digitized and measured in the frame of Archives of Photographic PLates for Astronomical USE (APPLAUSE) project (Tuvikene et al. 2014). We downloaded from its Data Release 3 archive[12] the measurements calibrated to "plate natural system" which is defined as

$$\mathrm{natmag} = V + \mathrm{color\_term} \cdot (B - V) \qquad (4)$$

[12] APPLAUSE archive is available at https://www.plate-archive.org/.

with color_term being fit by APPLAUSE pipeline for every individual frame, and *B* and *V* magnitudes corresponding to the Fourth U.S. Naval Observatory CCD Astrograph Catalog (UCAC4, Zacharias et al. (2013)) values. The latter is based on photometry from APASS DR6, and has zero point differences of 0.07 and 0.08 in *B* and *V* bands relative to APASS DR10 in the region around MN 112.

Assuming that brightness variations of the object are statistically independent from observing conditions (thus, the colour term) we may independently regress for the best *B* − *V* colour in Eq. 4 that minimizes the *V* band light curve spread. For the data for MN 112 available in APPLAUSE it gives

$$(B - V)_{\mathrm{APPLAUSE}} = 2.3 \pm 0.1 \,. \qquad (5)$$

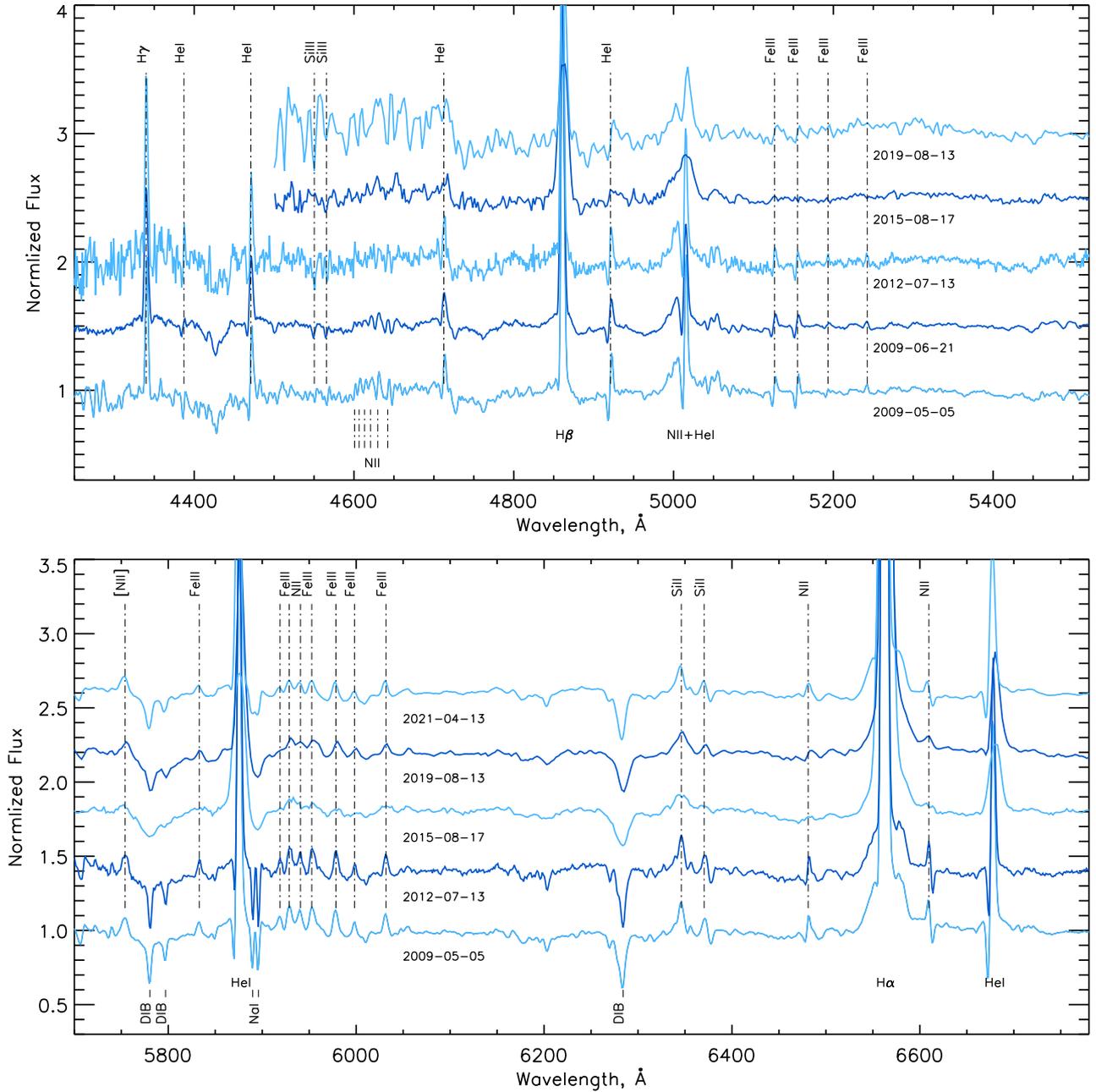

**Figure 3.** Spectra of MN 112 normalized by the local continuum level and vertically shifted for clarity. The spectra are acquired with different spectral resolutions (see Table 1), and some fine details of spectral lines (e.g. P Cyg profiles of He I lines, and complex structure of diffuse interstellar bands (DIBs)) are not apparent in the lower resolution ones.

We will use this colour for reducing APPLAUSE measurements to UCAC4 *V* band measurements, that may then be converted to APASS_*V* ones by subtracting 0.08 zero point offset, and then – to Pan-STARRS *g* magnitudes using Eq. 2.

### 2.2.7 Additional data

We used the imaging data acquired with SCORPIO instrument on Russian 6-m telescope during our spectral observations described in Section 2.1, along with historical images available for MN 112 in SCORPIO archive. We also found several images of the object in the public data archive of 2.5 m Isaac Newton Telescope (INT)[13] We processed all these images using the same pipeline as described in Section 2.2.2, calibrating the photometry to Pan-STARRS DR1 field stars. For SCORPIO, we used the photometric transformations between Pan-STARRS and Johnson-Cousins systems derived by Kostov & Bonev (2018). We then used the colour from Eq. 1 to reduce the measurements in red filters to *g* band.

The measurements performed in *g* and *r* filters on INT on the same

---

[13] Data are available in Isaac Newton Group Archive at http://casu.ast.cam.ac.uk/casuadc/ingarch/query.

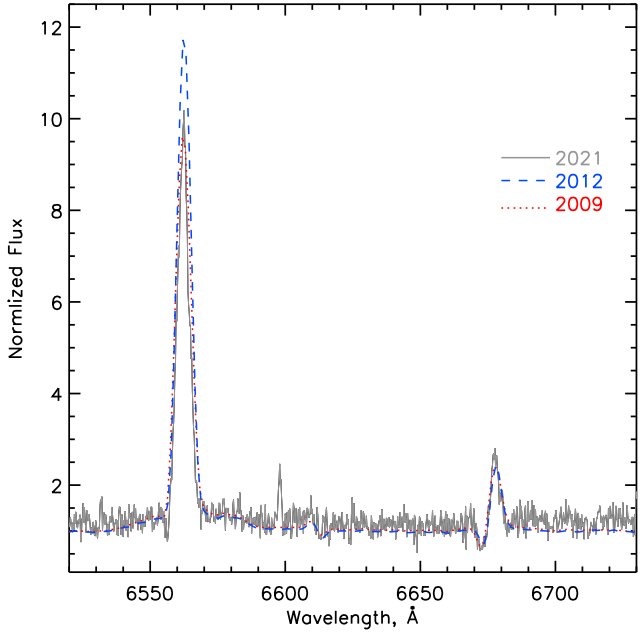

**Figure 4.** H$\alpha$ region spectra with medium and low resolution acquired with Calar-Alto-TWIN and Ondřejov observatory spectrographs in different years.

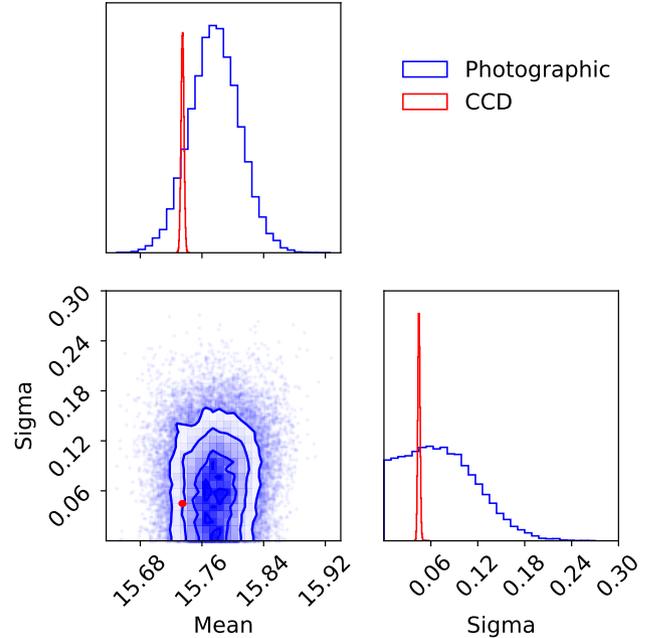

**Figure 5.** Posterior distributions of the photographic (blue) and CCD-based (red) data sets of the light curve shown in Figure 2, modeled using a Gaussian model with Gaussian intrinsic scatter to represent physical variability of the object.

night gives the colour

$$g - r = 2.34 \pm 0.02 \quad (6)$$

which is consistent with Eq. 1, while simultaneous observations on Russian 6-m give the mean colour

$$(B - V)_{\rm SCORPIO} = 2.62 \pm 0.02 \quad (7)$$

which unchanging between individual nights and is also consistent, within error bars, with APASS one from Eq. 3.

We also adapted the photometric estimates published in Paper I and based on their dedicated observations on a 40-cm Meade telescope of Cerro Armazones Astronomical Observatory of the Northern Catholic University (Antofagasta, Chile), as well as their re-calibration of the data from POSS-II B plates and the Guide Star Catalog 2.2 (McLean et al. 2000). We used their estimation of colour

$$(B - V)_{\rm Gvaramadze} = 2.60 \pm 0.12 \quad (8)$$

to reduce these values to green band, and then the transformations from Kostov & Bonev (2018) – to get Pan-STARRS $g$ band brightness estimations.

Paper I also provided the mean brightness estimate of $B = 17.2 \pm 0.2$ made on the number of photographic plates from Sternberg Astronomical Institute (Moscow, Russia) plate collection. We combined this value with the dates of individual observations to complement the light curve, converting the brightness to $g$ band in the same way as described above.

We also initiated a series of photometric observations of MN 112 on FRAM-ORM, which is a 10-inch Meade f/6.3 Schmidt-Cassegrain telescope with custom Moravian Instruments G2 CCD installed in Roque de Los Muchachos observatory, La Palma. The data were acquired in Johnson-Cousins $B$, $V$ and $R$ filters, and photometry was calibrated to Pan-STARRS DR1 field stars using the transformations from Kostov & Bonev (2018).

## 3 RESULTS

### 3.1 Spectral and photometric variability

Though the data collected in Section 2.1 have different spectral resolutions, we clearly see that during last 12 years since beginning of monitoring the spectrum of MN 112 did not change (Figure 3). The only difference which we may detect is the difference in intensities of H$\alpha$ lines in Calar-Alto TWIN spectra acquired in 2009 and 2012 (Figure 4). Such variability is typical for wind emission lines in spectra of blue supergiants (e. g. Clark et al. (2012); Gvaramadze et al. (2018)).

The IR spectrum shown in Figures 8 and 9 displays a prominent complex structure around hydrogen lines. This structure may either be interpreted as two-peaked hydrogen lines, suggesting a fast rotation like seen in Be stars, or alternatively as a combination of hydrogen and helium emission lines. In order to clarify it, we acquired a medium resolution spectrum of H$\alpha$ region in Ondřejov observatory. It allowed us to directly investigate the detailed shape of H$\alpha$ that lacks nearby He I lines. The spectrum shown in Figure 4 clearly displays a simple single-peak structure of H$\alpha$, thus rejecting the former variant. Therefore we may conclude that the complex structure seen in IR data is a combination of helium and hydrogen lines.

The light curve assembled from the diverse photometric data described in Section 2.2 and covering nearly 100 years since 1926 is shown in Figure 2, and summarized in Table 2. The long-term photometric data do not show any signs of large-amplitude variability or eruptions, while well-sampled and better quality light curve covering last 7 years display some small-amplitude variations on time scales from days to one year which, as Figure 1 shows, keep the $g - r$ colour essentially stable, down to about 0.01 magnitude.

In order to characterize the amplitude of the light curve variations we separately modeled both the photographic (pre-1995, lower accuracy) and CCD-based (since 2009, better quality) data using a Gaussian model with intrinsic scatter, with the latter also represented by a Gaussian and reflecting the variability of the object. The resulting posterior distributions of the mean value and standard deviation of intrinsic light curve are shown in Figure 5, and are essentially consistent between both data sets. For the CCD data set, the amplitude of intrinsic variations is $\sigma = 0.045^{+0.002}_{-0.002}$ (with 90% confidence), while for photographic it is $\sigma = 0.069^{+0.084}_{-0.062}$ (with 90% confidence). It means that over the last century the amplitude of light curve variations did not exceed 0.15 magnitudes.

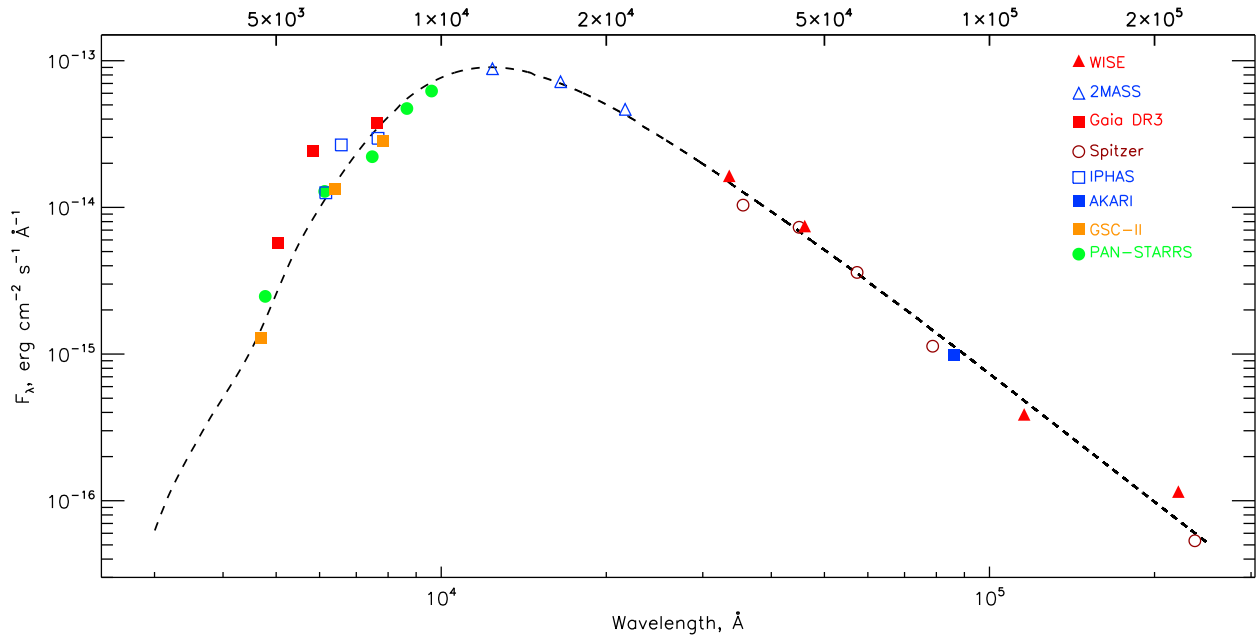

**Figure 6.** Observed spectral energy distribution of MN 112 based on the multi-wavelength photometric data as described in Section 3.2, compared to the continuum of the reddened model spectrum. The model spectrum is scaled for the distance d=13.53 kpc and the interstellar extinction ($E(B-V) = 2.65$) is applied to it.

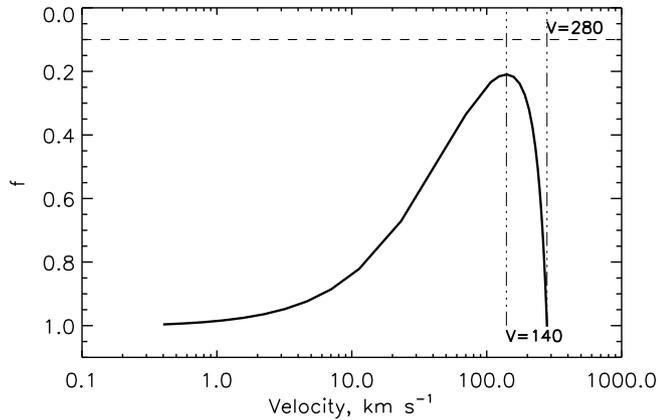

**Figure 7.** Profile of clumping $f(r)$ in the stellar wind, expressed as a function of wind radial velocity $v(r)$, as described in Section 3.2

**Table 3.** Stellar parameters of MN 112 estimated for two distances (see the text for details).

|  | Kostenkov et al. (2020) | This work |
|---|---|---|
| $d$ (kpc) | 6.93 | 13.53 |
| $E(B-V)$ | 2.72[a] | 2.65 |
| $L_*$ ($\times 10^5 L_\odot$) | 5.77 | 15 |
| $\log L_*/L_\odot$ | 5.76 | 6.18 |
| $\dot{M}$ ($\times 10^{-5} M_\odot$ yr$^{-1}$) | 1.815 | $6.8 \pm 0.3$ |
| $T_*$ (kK) | 22.8 | $18.0 \pm 0.5$ |
| $R_*$ ($R_\odot$) | 49 | $126 \pm 7$ |
| $T_{\rm eff}$ (kK) | $15.2 \pm 0.5$ | $17.4 \pm 0.5$ |
| $R_{2/3}$ ($R_\odot$) | 110 | 135 |
| $V_\infty$ (km s$^{-1}$) | $300 \pm 10$ | 280 |
| $\beta$ | $3.4 \pm 0.2$ | 1.0 |
| $f$ | $0.10 \pm 0.05$ | 0.1 |

[a] Kostenkov et al. (2020) used $A_V = 8.45 \pm 0.11$

### 3.2 Spectral modelling

The first modeling of the MN 112 atmosphere was performed by Kostenkov et al. (2020), who used non-LTE radiative transfer code CMFGEN (Hillier & Miller 1998) for the analysis. To determine the luminosity of the object the authors applied a geometric distance $d = 6.93^{+2.74}_{-1.81}$ kpc taken from the second *Gaia* Data Release (DR2; Bailer-Jones et al. (2018)). However according to the *Gaia* DR3 the distance to MN 112 is $d = 13.53^{+2.64}_{-2.07}$ kpc (Bailer-Jones et al. 2021), which significantly exceeds previous estimates. Thus, we recalculated the model of MN 112 atmosphere based of new measure of distance and spectra covering both optical and IR ranges.

To determine the stellar parameters we have used CMFGEN in the same way as we did it early in (for example Gvaramadze et al. (2019); Maryeva et al. (2020)). Our modeling included the following major steps:

- measurement of effective temperatures;
- measurement of interstellar reddening $E(B-V)$;

- measurement of luminosity;
- measurement of other parameters.

To measure the effective temperatures $T_{\rm eff}$[14] we used the intensities of Si II – Si III lines and N II, Fe III, as well as presence of He I lines, and absence of He II, N III and Fe II.

The colour excess $E(B-V)$ towards MN 112 was estimated by comparing the spectral energy distribution (SED) in the model spectrum with the observed photometric measurements (Figure 6) taken from the following sky surveys: WISE (Cutri et al. 2021); Two Micron All Sky Survey (2MASS) (Cutri et al. 2003); *Gaia* eDR3 (Gaia Collaboration 2020); the Galactic Legacy Infrared Midplane Survey Extraordinaire (GLIMPSE) Source Catalog (Spitzer Science 2009); Photometric H-Alpha Survey of the Northern Galactic Plane (IPHAS) (Barentsen et al. 2014); *AKARI* Far-infrared All-Sky Survey data (Ishihara et al. 2010; Doi et al. 2015); Guide Star Catalog II (Lasker et al. 2008); the Panoramic Survey Telescope and Rapid Response System (Pan-STARRS) DR1 (Chambers et al. 2016).

The stellar wind was assumed to be clumpy with an empty interclump medium (Hillier & Miller 1999). The wind volume filling factor $f = \bar{\rho}/\rho(r)$, where $\bar{\rho}$ is the homogeneous (unclumped) wind density and $\rho(r)$ is the density in clumps (assumed to be optically thin), depends on radius as $f(r) = f_\infty + (1 - f_\infty) \exp(-v(r)/v_{cl})$, where $f_\infty$ characterizes the density contrast and $v_{cl}$ is the velocity at which clumping starts. Davies et al. (2005) based on spectropolarimetric observations of LBVs demonstrated that the clumping starts in the innermost layers of their winds; therefore, in our modeling we adopted $v_{cl}$=50 km s$^{-1}$ and $f_\infty = 0.1$. Clumping in dense winds of hot stars decreases outwards (Puls et al. 2006; Najarro et al. 2011). Based on that in our computations we assumed that the clumping has to start to disappear at the distance where the wind velocity exceeds 140 km s$^{-1}$ (Figure 7) for simultaneous fitting of the visible and near-infrared parts of the spectrum.

CMFGEN uses $\beta$-velocity law approximation for describing the radial dependence of the wind velocity. Velocity profile in winds of hot stars is complex non-monotonous function (see for example Krtička & Kubát (2017); Sander et al. (2017)). However hydrodynamical calculations by Gormaz-Matamala et al. (2021) show that the shape of velocity law does not effect on measurements of basic parameters such as luminosity and temperature. As the main aim of our modeling is determination of position of the star on H–R diagram, in our calculations we used simple $\beta$-velocity law with $\beta = 1.0$.

Figures 8 and 9 present final best-fit model, which describes all basic spectral details in the 380–2470 nm wavelength range. The derived parameters of MN 112 are listed in Table 3 which also for comparison contains the parameters estimated by Kostenkov et al. (2020). As expected, our measurement of luminosity is significantly larger than the one by Kostenkov et al. (2020). This difference in luminosity as well as difference in used wind velocity law lead to the significant difference in the mass loss rate.

## 4 DISCUSSION

Our study covered 12 years of spectral observations and nearly 100 years of photometric monitoring of MN 112. Based on these data we reliably registered only low amplitude photometric variability, with variations of about 0.05 magnitudes in Pan-STARRS $g$ band on time scales from days to about a year, and with $g - r$ colour remaining constant. There are no signs of any significant variability with amplitude exceeding 0.15 magnitudes on longer time scales.

---

[14] $T_*$ and $T_{\rm eff}$ are effective temperatures at the bottom of the wind and at radius $R_{2/3}$ where Rosseland optical depth is 2/3, correspondingly.

**Table 4.** Abundances of main chemical elements in the atmospheres of MN 112 and AG Car.

|   | MN 112 | AG Car |
|---|---|---|
| H | $0.37 \pm 0.03$ | 0.36 |
| He | $0.62 \pm 0.03$ | 0.62 |
| N | $(4.0 - 8.0) \times 10^{-3}$ | $7.2 \times 10^{-3}$ |
| Published | This work | Groh et al. (2009a) |

Such small-amplitude variability with ⩽0.1 mag variations on a timescales of days to weeks is typical for luminous stars of early types (the '$\alpha$ Cygni' variables; e.g.van Genderen et al. (1998); van Leeuwen et al. (1998); Clark et al. (2010)), and is likely related to heat- or convection-driven nonradial waves (Jiang et al. 2018; Elliott et al. 2022) that are excited in the stellar envelope. It is also detected in the majority of known LBVs (Abolmasov 2011; Dorn-Wallenstein et al. 2019; Grassitelli et al. 2021).

In Paper I based on the spatial closeness of the object and estimation of its bolometric magnitude it has been suggested that the star belongs to NGC 6823 cluster (which is s part of Vul OB1 association) in local (Orion) spiral arm. However, new distance estimates force us to reconsider the position of the star in the Galaxy. According to this new distance MN 112 could be a member of Outer Scutum-Centaurus arm (OSC) (Armentrout et al. 2017; Wenger et al. 2018), which is the most distant molecular spiral arm known in the Milky Way (Figure 10). H I regions located in OSC show radial velocities around -80 km s$^{-1}$(Armentrout et al. 2017). Unfortunately, there are no absorption lines in the spectrum of MN 112 that may be used for measuring its radial velocity and testing this hypothesis. And while the distance estimate may in principle still be modified in upcoming *Gaia* data releases, we should note that MN 112 cannot possibly lay even farther away as it should be associated with some Galactic spiral arm where all massive stars form. Thus, current *Gaia* DR3 based distance may be considered an upper limit, and the luminosity (and therefore mass) of MN 112 estimated from it – also largest possible one.

According to our results of modeling and new distance estimate MN 112 moved to upper part of H–R diagram and is now laying on the evolutionary track for initial mass of 70 M$_\odot$ (Figure 11), close to famous LBV AG Car. Table 4 shows that both stars have similar abundances of hydrogen, helium and nitrogen, that suggests that their evolutionary status is also the same. In this context, the lack of S Dor type variability in MN 112 looks even more surprising.

On the other hand, current position of MN 112 in H–R diagram is also close to Schulte 12 hypergiant, which is one of the most luminous stars in the Galaxy located above Humphreys-Davidson instability limit, and at the same time not behaving as an LBV (Clark et al. 2012). However, unlike MN 112, Schulte 12 has unevolved chemical composition (Clark et al. 2012). Therefore we cannot conclude that MN 112 is also just a blue hypergiant which has not yet reached the phase of strong variability. Enriched abundances of helium and nitrogen rather indicate that MN 112 already evolved past LBV activity phase.

In comparison with Paper 1 we examined the structure of circumstellar nebula of MN 112 in more details and found that it has a complex shape: the star is surrounded by extended circular nebula with diameter ≈ 1.9′ and smaller nebula with rectangular shape with longer side $\approx 0.9'$ (Figure 12). Such complex structure may point towards some historical change in the stellar wind properties – a spherically-symmetric outer shell was produced by a slow wind, while bipolar inner one is the result of fast wind accelerated by rotation of the star, or interaction with the second component in the binary system. Such bipolar structures are observed e.g. around AG Car ($1.4\times 2$ pc) or HR Car ($0.65\times 1.3$ pc), although with smaller sizes. The size of circumstellar nebula of MN 112 supports the hypothesis about post-LBV status of the star. Its linear sizes (radius of spherical part is $\approx 3.7$ pc, size of bipolar part is $\approx 3.55$ pc) are much larger than the typical ones for LBV nebulae ($< 1$ pc, Weis & Bomans (2020)) – on the other hand, MN 112's nebula is similar in size to nebulae around Wolf–Rayet stars (e.g. Stock & Barlow (2010); Toalá et al. (2015)). As forbidden sulphur lines [SII] $\lambda$6716, 6731 are not detected in the spectrum of MN 112, we could not estimate electron density of the nebula, and therefore its mass.

**Figure 8.** The normalized spectra of MN 112 (grey thick line) compared with the best-fit CMFGEN model (blue thin line) with the parameters as given in Table 3. Observed spectra obtained with (from the top panel downwards): Calar-Alto+TWIN (July, 2012); Russian 6-m+SCORPIO (April, 2021); IRTF+SpeX (June, 2010).

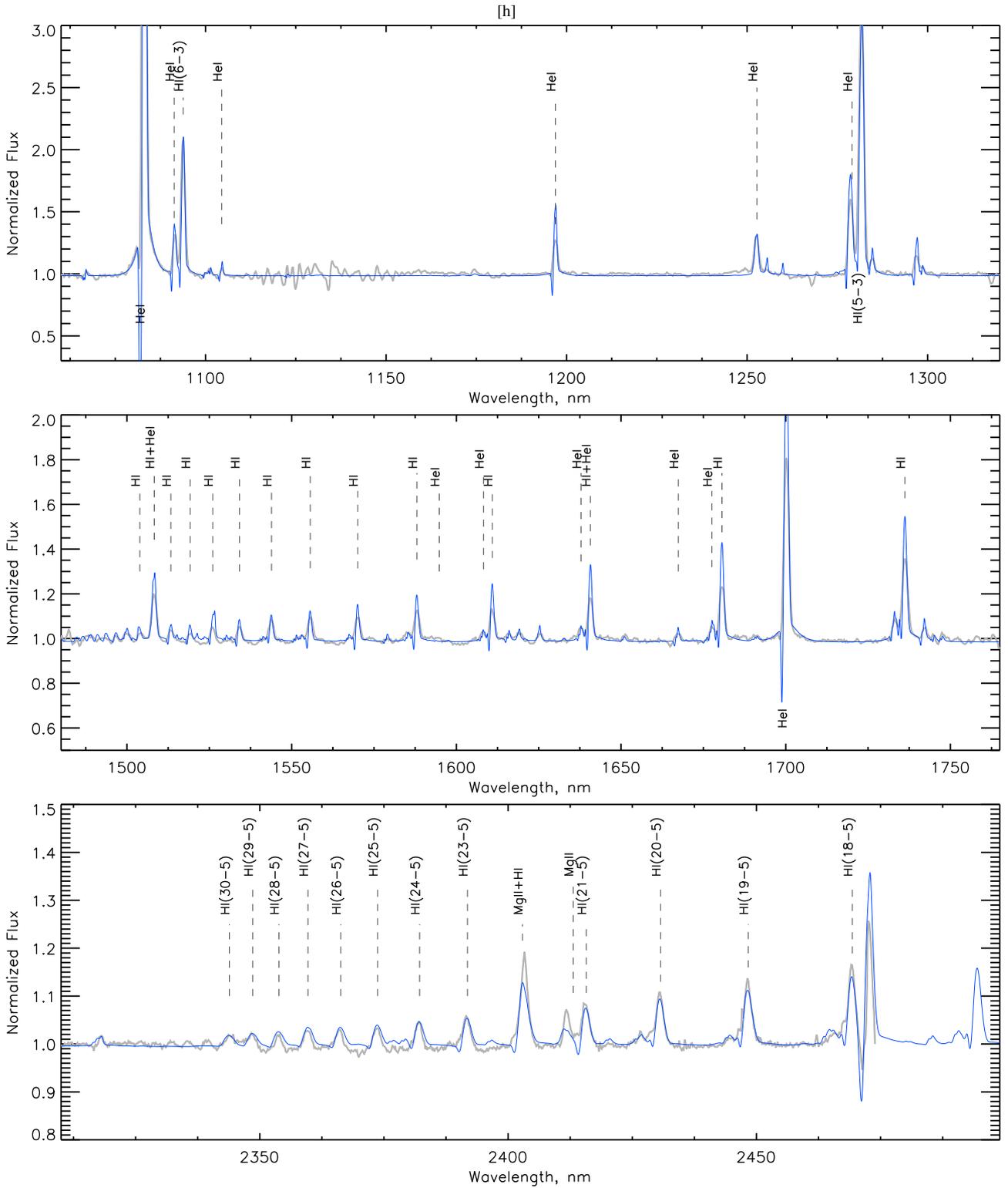

**Figure 9.** The normalized spectrum of MN112 (grey thick line) compared with the best-fit CMFGEN model (blue thin line) with the parameters as given in Table 3. Two upper spectra obtained with IRTF+SpeX (June, 2010) and the bottom spectrum obtained with Gemini+GNIRS (July 2013).

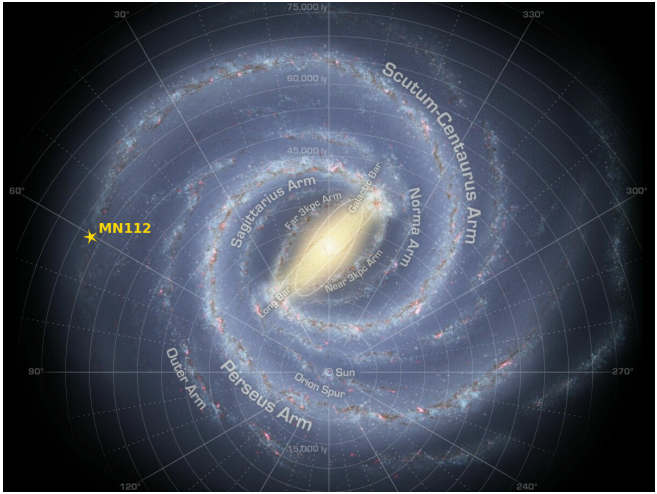

**Figure 10.** Schematic representation of MN 112's location in the Galaxy. Image credit: NASA/Adler/U. Chicago/Wesleyan/JPL-Caltech

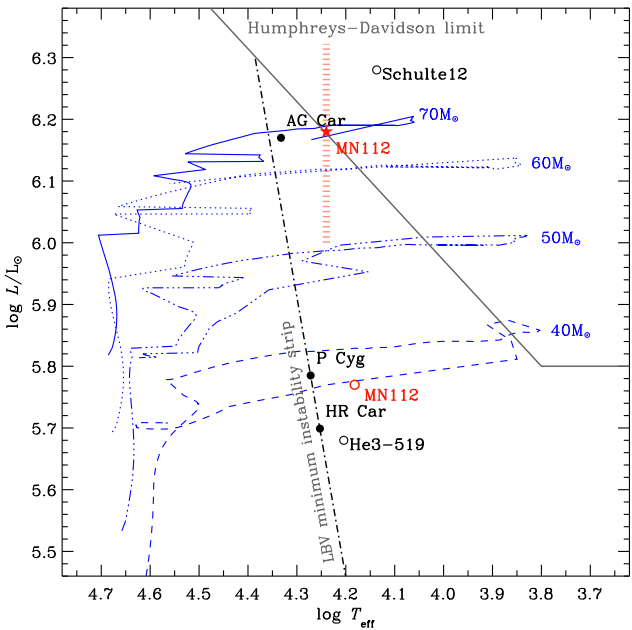

**Figure 11.** Position of MN 112 in the H–R diagram according to the calculations by Kostenkov et al. (2020) (red circle) and according to our new estimations (red star). Blue horizontal lines show Geneva evolutionary tracks (Ekström et al. 2012) for star with mass 35, 40, 50 and 70 $M_\odot$ in solar metallicity with initial rotational velocity 40 per cent of breakup. In addition, the positions of LBV stars AG Car (Groh et al. 2009a), P Cygni (Najarro 2001) and HR Car (Groh et al. 2009b) are shown with filled circles, cLBV Schulte 12 (Clark et al. 2012) and He 3-519 (Crowther et al. 1997) with unfilled circle. Grey solid line is Humphreys-Davidson limit (Humphreys & Davidson 1994), grey dash-dotted line is LBV minimum instability strip as defined in Groh et al. (2009b).

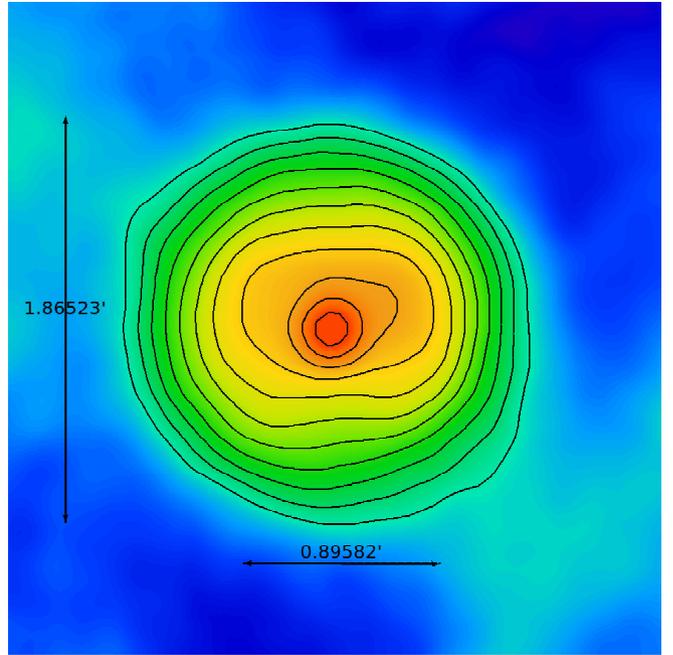

**Figure 12.** WISE $22\mu$m image of MN 112. Vertical arrow shows size of spherically symmetrical part of nebula, while horizontal one – size of bipolar internal structure. Black contours represent the isophotes of intensity distribution.

Main characteristic of LBVs is S Dor instability. Its physical drivers are still poorly understood, but probably it is related to radiation pressure instability (Humphreys & Davidson 1994; Vink 2012), pulsations (Lovekin & Guzik 2014), or is the result of wind-envelope interaction (Grassitelli et al. 2021). Thus, duration of LBV phase when the star displays this kind of activity is one of open questions in the field of massive stars. Using stellar evolutionary tracks and atmospheric modeling Groh et al. (2014) showed that stars with 60 $M_\odot$ spend $2 \times 10^5$ yr in the LBV phase, only 5% of their lifetime. Kalari et al. (2018) using statistical census of S Dor variables among the population of blue supergiants in Small Magellanic Cloud estimated maximum duration of the LBV phase as a few $10^3$ years. Also important is how exactly the LBV phase look like – do LBVs show S Dor variability during all the phase, or are they switching between activity and quiescence intervals? S Dor instability is detected for almost every *bona fide* LBV except for P Cyg, which is, however, known to show at least two giant eruptions in the past (in 1600 and in 1655), and presently displays just a low-amplitude variability (Humphreys & Davidson 1994; de Groot et al. 2001; Elliott et al. 2022). Long dormant period of at least 50 years was also seen in the century-long light curve of extragalactic LBV Romano's star (Polcaro et al. 2016; Maryeva et al. 2019) prior to its recent outbursts. Another example of such dormant LBV was MCA-1B that also recently showed an outburst activity after at least 20 years of sleep (Smith et al. 2020).

Collected spectral and photometric data argue that over past century MN 112 does not exhibit S Dor variability. However, its other characteristics – position in H–R diagram, evolved chemical composition, slow wind, presence of circumstellar nebula – are consistent with MN 112 being LBV. Therefore we suggest to classify this object as a dormant LBV. We suggest that detailed study of such objects would allow us to improve evolutionary models for evolved massive stars, and expand our understanding of how long the quiescent phases of LBV stars may last – for how long LBVs may "sleep".

## 5 CONCLUSIONS

In this paper we studied MN 112 which were earlier classified as LBV candidate (Paper 1). We combined photometric data covering almost 100 years and spectral data covering 12 last years. We detected only low-amplitude variability and variability of intensity of H$\alpha$ line, which is typical for both LBVs and blue supergiants. According to the results of numerical modeling the star is located in upper part of H–R diagram near AG Car and its initial mass was around 70 M$_\odot$. Despite the lack of characteristic S Dor variability, based on overall characteristics of the star and its position in H–R diagram we suggest that it still may be considered LBV candidate – that MN 112 is a dormant LBV that may be used for studying long-term behaviour of that "sleeping" phase of these evolved massive stars.

**ACKNOWLEDGMENTS**

We dedicate this publication to the memory of our dear colleague Vasilii Gvaramadze, who passed away on September 2nd, 2021. The search for evolved stellar objects by means of detection of their circumstellar nebulae in modern sky surveys has been the topic that Vasilii successfully developed for last ten years, and he assembled and inspired our diverse team with his intellectual and organizational talents. This work has not been possible without his efforts.

We are grateful to Michaela Kraus for providing us K-band spectrum from GEMINI and optical one from Tartu 1.5-m telescope, to Tiina Liimets who obtained Ondřejov spectrum used in this study, Alexandra Zubareva who helped to clarify the dates of observations for Moscow photographic plates.

The work partially funded from the European Union's Framework Programme for Research and Innovation Horizon 2020 (2014-2020) under the Marie Skłodowska-Curie Grant Agreement No. 823734. A.Y.K. acknowledges support from the National Research Foundation (NRF) of South Africa. S.K. acknowledges support from the European Structural and Investment Fund and the Czech Ministry of Education, Youth and Sports (Project CoGraDS – CZ.02.1.01/0.0/0.0/15_003/0000437).

This work based on data taken with the Perek telescope at the Astronomical Institute of the Czech Academy of Sciences in Ondřejov, which is supported by the project RVO:67985815 of the Academy of Sciences of the Czech Republic. This study is partially based on the data obtained at the unique scientific facility the Big Telescope Alt-azimuthal of SAO RAS and was supported under the Ministry of Science and Higher Education of the Russian Federation grant 075-15-2022-262 (13.MNPMU.21.0003).

This work is partially based on archival data obtained with the NASA Infrared Telescope Facility, which is operated by the University of Hawaii under a contract with the National Aeronautics and Space Administration.

This publication makes use of data products from the Wide-field Infrared Survey Explorer, which is a joint project of the University of California, Los Angeles, and the Jet Propulsion Laboratory/California Institute of Technology, funded by the National Aeronautics and Space Administration.

The operation of the robotic telescope FRAM-ORM is supported by the grant of the Ministry of Education of the Czech Republic LM2018102. The data calibration and analysis related to the FRAM-ORM telescope are supported by the Ministry of Education of the Czech Republic MSMT-CR LTT18004, MSMT/EU funds CZ.02.1.01/0.0/0.0/16_013/0001402 and CZ.02.1.01/0.0/0.0/18_046/0016010.

This work is based in part on observations obtained with the Samuel Oschin 48-inch Telescope at the Palomar Observatory as part of the Zwicky Transient Facility (ZTF) project. ZTF is supported by the National Science Foundation under Grant No. AST-1440341 and a collaboration including Caltech, IPAC, the Weizmann Institute for Science, the Oskar Klein Center at Stockholm University, the University of Maryland, the University of Washington, Deutsches Elektronen-Synchrotron and Humboldt University, Los Alamos National Laboratories, the TANGO Consortium of Taiwan, the University of Wisconsin at Milwaukee, and Lawrence Berkeley National Laboratories. Operations are conducted by Caltech Optical Observatories (COO), the Infrared Processing and Analysis Center (IPAC), and the University of Washington (UW).

The work is partially based on the data from DASCH project at Harvard which is partially supported from NSF grants AST-0407380, AST-0909073, and AST-1313370. It is also uses the data from APPLAUSE project which has been funded by DFG (German Research Foundation, Grant), Leibniz Institute for Astrophysics Potsdam (AIP), Dr. Remeis Sternwarte Bamberg (University Nuernberg/Erlangen), the Hamburger Sternwarte (University of Hamburg) and Tartu Observatory.

This paper makes use of data obtained from the Isaac Newton Group Archive which is maintained as part of the CASU Astronomical Data Centre at the Institute of Astronomy, Cambridge.

This research was made possible through the use of the AAVSO Photometric All-Sky Survey (APASS), funded by the Robert Martin Ayers Sciences Fund and NSF AST-1412587.

It also uses the data from the Pan-STARRS1 Surveys (PS1) and the PS1 public science archive, whose have been made possible through contributions by the Institute for Astronomy, the University of Hawaii, the Pan-STARRS Project Office, the Max-Planck Society and its participating institutes, the Max Planck Institute for Astronomy, Heidelberg and the Max Planck Institute for Extraterrestrial Physics, Garching, The Johns Hopkins University, Durham University, the University of Edinburgh, the Queen's University Belfast, the Harvard-Smithsonian Center for Astrophysics, the Las Cumbres Observatory Global Telescope Network Incorporated, the National Central University of Taiwan, the Space Telescope Science Institute, the National Aeronautics and Space Administration under Grant No. NNX08AR22G issued through the Planetary Science Division of the NASA Science Mission Directorate, the National Science Foundation Grant No. AST-1238877, the University of Maryland, Eotvos

Lorand University (ELTE), the Los Alamos National Laboratory, and the Gordon and Betty Moore Foundation.

## DATA AVAILABILITY

The data underlying this article will be shared on reasonable request to the corresponding author.

This paper has been typeset from a T<sub>E</sub>X/L<sup>A</sup>T<sub>E</sub>X file prepared by the author.